\documentclass[final,5p]{elsarticle}

\usepackage{graphicx}

\usepackage{subfigure}

\usepackage{times}

\newcommand{\KK}{\ensuremath{K^+K^-}~}

\newcommand{\mmd}{\ensuremath{\mbox{MM}_d(\gamma,\phi)}}
\newcommand{\Eg}{\ensuremath{{E}_{\gamma}}~}

\newcommand{\tvarr}{\ensuremath{\tilde{t}}~}

\journal{Physics Letter B} 

\begin{document}

\begin{frontmatter}


\title{Measurement of the incoherent $\gamma d \rightarrow \phi p n$
photoproduction near threshold}


\author[a]{W.C.~Chang}
\author[b]{M.~Miyabe}
\author[c]{T.~Nakano}
\author[c,d]{D.S.~Ahn}
\author[d]{J.K.~Ahn}
\author[e]{H.~Akimune}
\author[f]{Y.~Asano}
\author[g]{S.~Dat\'{e}}
\author[c]{H.~Ejiri}
\author[h]{H.~Fujimura}
\author[c,s]{M.~Fujiwara}
\author[i]{S.~Fukui}
\author[c]{S.~Hasegawa}
\author[j]{K.~Hicks}
\author[c]{K.~Horie}
\author[c]{T.~Hotta}
\author[b]{K.~Imai}
\author[h]{T.~Ishikawa}
\author[k]{T.~Iwata}
\author[c]{Y.~Kato}
\author[l]{H.~Kawai}
\author[c]{K.~Kino}
\author[c]{H.~Kohri}
\author[g]{N.~Kumagai}
\author[m]{S.~Makino}
\author[n]{T.~Matsuda}
\author[o]{T.~Matsumura}
\author[c]{N.~Matsuoka}
\author[c]{T.~Mibe}
\author[p]{M.~Miyachi}
\author[c,s]{N.~Muramatsu}
\author[b]{M.~Niiyama}
\author[q]{M.~Nomachi}
\author[g]{Y.~Ohashi}
\author[g]{H.~Ohkuma}
\author[l]{T.~Ooba}
\author[a]{D.S.~Oshuev}
\author[r]{C.~Rangacharyulu}
\author[q]{A.~Sakaguchi}
\author[w]{P.M.~Shagin}
\author[l]{Y.~Shiino}
\author[h]{H.~Shimizu}
\author[q]{Y.~Sugaya}
\author[c]{M.~Sumihama}
\author[n]{Y.~Toi}
\author[g]{H.~Toyokawa}
\author[x]{M.~Uchida}
\author[t]{A.~Wakai}
\author[a]{C.W.~Wang}
\author[a]{S.C.~Wang}
\author[e]{K.~Yonehara}
\author[c,g]{T.~Yorita}
\author[u]{M.~Yoshimura}
\author[c]{M.~Yosoi}
\author[v]{R.G.T.~Zegers}
\author{(LEPS Collaboration)}

\address[a]{Institute of Physics, Academia Sinica, Taipei 11529, Taiwan}
\address[b]{Department of Physics, Kyoto University, Kyoto 606-8502, Japan}
\address[c]{Research Center for Nuclear Physics, Osaka University, Ibaraki, Osaka 567-0047, Japan}
\address[d]{Department of Physics, Pusan National University, Busan 609-735, Republic of Korea}
\address[e]{Department of Physics, Konan University, Kobe, Hyogo 658-8501, Japan}
\address[f]{XFEL Project Head Office, RIKEN 1-1, Koto Sayo Hyogo 679-5148, Japan}
\address[g]{Japan Synchrotron Radiation Research Institute, Sayo, Hyogo 679-5143, Japan}
\address[h]{Laboratory of Nuclear Science, Tohoku University, Sendai, Miyagi 982-0826, Japan}
\address[s]{Advanced Photon Research Center, Japan Atomic Energy Agency, Kizugawa, Kyoto 619-0215, Japan}
\address[i]{Department of Physics and Astrophysics, Nagoya University, Nagoya, Aichi 464-8602, Japan}
\address[j]{Department of Physics And Astronomy, Ohio University, Athens, Ohio 45701, USA}
\address[k]{Department of Physics, Yamagata University, Yamagata 990-8560, Japan}
\address[l]{Department of Physics, Chiba University, Chiba 263-8522, Japan}
\address[m]{Wakayama Medical College, Wakayama, Wakayama 641-8509, Japan}
\address[n]{Department of Applied Physics, Miyazaki University, Miyazaki 889-2192, Japan}
\address[o]{Department of Applied Physics, National Defense Academy in Japan, Yokosuka, Kanagawa 239-8686, Japan}
\address[p]{Department of Physics, Tokyo Institute of Technology, Tokyo 152-8551, Japan} 
\address[q]{Department of Physics, Osaka University, Toyonaka, Osaka 560-0043, Japan}
\address[r]{Department of Physics and Engineering Physics, University of Saskatchewan, Saskatoon SK S7N 5E2, Canada}
\address[w]{School of Physics and Astronomy, University of Minnesota, Minneapolis, Minnesota 55455, USA}
\address[x]{Department of Physics, Tokyo Institute of Technology, Tokyo 152-8551, Japan}
\address[t]{Akita Research Institute of Brain and Blood Vessels, Akita 010-0874, Japan}
\address[u]{Institute for Protein Research, Osaka University, Suita, Osaka 565-0871, Japan}
\address[v]{National Superconducting Cyclotron Laboratory, Michigan State University, East Lansing, Michigan 48824, USA}


\begin{abstract}

We report measurements of differential cross sections and decay
asymmetries of incoherent $\phi$-meson photoproduction from the
deuteron at forward angles using linearly polarized photons at
\Eg=1.5-2.4 GeV. The nuclear transparency ratio for the deuteron
shows a large suppression, and is consistent with the A-dependence of
the ratio observed in a previous measurement with nuclear targets. The
reduction for the deuteron cannot be adequately explained in term of
isospin asymmetry. The present results suggest the need of refining
our understanding of the $\phi$-N interaction within a nucleus.

\end{abstract}



\end{frontmatter}

\section{Introduction}

The origin of hadron mass is one of the most fundamental questions in
understanding the strong interaction. It is generally believed that
the dynamical breaking of chiral symmetry is responsible for formation
of composite hadrons and the origin of their masses. A partial
restoration of chiral symmetry~\cite{chiral} may occur in hot or dense
nuclear matter. The corresponding change of the chiral condensate
would lead to a modification of hadron properties in the nuclear
medium. There have been many experiments to search for such
modifications in the collisions of hadron, heavy-ion and photon beams
with nuclear targets.

In particular the $\phi$ vector meson is a good probe because of its
narrow width in free space (4.26 MeV/$c^2$) and the small $\phi$-N
interaction cross section. In the case of the $\rho$ and $\omega$
mesons, the overlap of two peaks with different widths along with
strong final-state interactions with the nuclear medium, makes it
difficult to single out the intrinsic in-medium modification of either
one of the two mesons. In contrast, any modification to the lineshape
of $\phi$ meson in the nuclear medium remains mostly unaffected by
hadronic interactions, and can be identified more easily and with less
ambiguity. The first evidence for the in-medium mass shift of $\phi$
mesons was reported in the pA reaction at the normal nuclear
density~\cite{E325,Hayano}.

An alternative way is to determine the in-medium width of $\phi$
mesons by measuring the nuclear transparency ratio (or survival
probability) $T_{A}=\sigma_{A}/(A\sigma_{N})$, where $\sigma_{A}$ is
the incoherent $\phi$-meson production cross section from the nuclear
target with an atomic mass A, and $\sigma_{N}$ is the cross section
from a free proton~\cite{oset,phiA_oset}. A broadened width in the
nuclear medium would produce an increase of the $\phi$-N inelastic
cross section, thus leading to a drop of $T_{A}$ below
unity. Experimentally, strong attenuation was observed in the $\phi$
photoproduction from Li, C, Al and Cu nuclei~\cite{LEPS_phiA}. In a
simple Glauber approximation, the inelastic $\phi$-N cross section
$\sigma_{\phi N}^{\rm inel}$ was determined to be about 35 mb from the
measured A-dependence, which is significantly larger than the total
$\phi$-N cross section of about 13 mb estimated by using the
vector-meson dominance model in $\phi$ photoproduction off the proton
at $E_{\gamma}=$ 3-6 GeV~\cite{phip_DESY}. A recent measurement of
coherent $\phi$-meson photoproduction off the deuteron at CLAS is
compatible with a $\sigma_{\phi N}$ of 30 mb together with a large
transverse slope for the $\phi N \rightarrow \phi N$
process~\cite{CLAS_phiD_co}.

Subsequent theoretical studies~\cite{phiA_muhlich,phiA_sibirtsev} have
confirmed that it is necessary to introduce a large $\sigma_{\phi
N}^{\rm inel}$ in the range of 30-60 mb or invoke a coupled-channel
effect of $\omega$-$\phi$ mixing~\cite{phiA_sibirtsev} for describing
the measured A-dependence of attenuation in $\phi$-meson
photoproduction from nuclei. Note that one of these studies takes into
account the standard nuclear structure effects, such as Fermi motion,
Pauli blocking, nuclear shadowing and quasi elastic scattering
processes~\cite{phiA_muhlich}. The enlargement of $\sigma_{\phi
N}^{\rm inel}$ may be due to nuclear density effect in nuclear
matter~\cite{phiA_oset}. To clarify the underlying physics, it is
essential for us to construct a reliable baseline in a reasonably low
density region, using the simplest nucleus--the deuteron--made of one
proton and one neutron, where nuclear density effects are minimized.

In the high photon energy region of 45-85 GeV, the $\phi$-meson
photoproduction yields per nucleon for proton and deuteron targets
were found to be the same within a 10\%
discrepancy~\cite{Busenitz}. The $\phi$ mesons were produced with
large momenta and the results were well described using vector-meson
dominance and the additive quark model. In the near-threshold region,
the quark exchange processes, other than Pomeron exchange, could
contribute, and an interference among them would generate an isospin
dependence~\cite{Titov_iso}. In this Letter we report the differential
cross sections and decay asymmetries of incoherent
$\phi$-photoproduction with a liquid deuterium target at
$E_{\gamma}=$ 1.5-2.4 GeV at forward scattering angles.

\section{Experiment and analysis}


The experiment was carried out at the laser-electron photon facility,
SPring-8. Linearly polarized photons were generated by backward
Compton scattering of polarized laser light from 8 GeV electrons in
the storage ring. The high polarization of photon beams allowed a
measurement of the decay asymmetry. Charged particles emitted from the
interaction points of photon with a 16-cm long deuterium target were
detected at forward angles in the LEPS spectrometer. Particle
identification was made by the mass reconstruction using the time of
flight and momentum.  For more details on the detectors and the
quality of the particle identification, see Ref.~\cite{LEPS}.

\begin{figure}
\vspace{-1cm}
\includegraphics[width=0.55\textwidth]{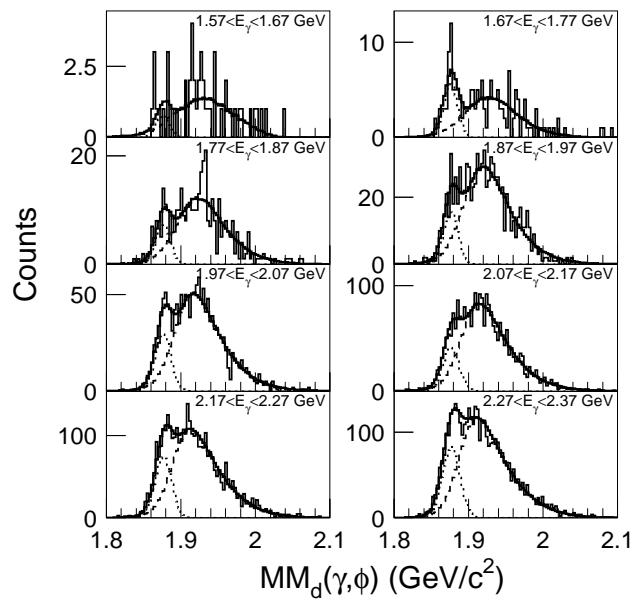}

\caption{The missing mass spectra assuming the whole deuteron as
target, $\mmd$, at $1.57<E_{\gamma}<2.37$ GeV. Each $\mmd$
spectrum is fitted with the sum (solid line) of MC-simulated
components of coherent (dotted line) and incoherent (dashed line)
events.}

\label{fig:mmdld2}
\end{figure}

Events of $\phi$-meson production were selected with a cut on the
invariant mass of a \KK pair $|{\rm M}(K^+K^-)-{\rm M}_{\phi}|< 0.01$
GeV/c$^2$. In the missing mass spectra, assuming the whole deuteron as
the target ($\mmd$), events of coherent $\phi$ production peak at the
deuteron mass of 1.875 GeV/c$^2$ whereas incoherent events, $\gamma
d \rightarrow \phi p n$, are distributed at relatively higher mass.
Figure~\ref{fig:mmdld2} shows the spectra of $\mmd$ in eight bins of
measured photon energy range of $1.57<E_{\gamma}<2.37$ GeV. A
narrow structure is seen at $\mmd=1.92$ (GeV/$c^2$) in the bin of
$1.77<E_{\gamma}<1.87$ GeV. Because its statistical significance is
less than 2$\sigma$ and no similar structure is observed at the same
missing mass region in the other $E_{\gamma}$ bins, it is interpreted
as due to statistical fluctuation.

The missing-mass spectra are nicely reproduced by the Monte Carlo (MC)
simulated distributions of coherent and incoherent $\phi$ production
processes. Using the GEANT3 software package~\cite{GEANT}, MC
simulation took into account experimental parameters such as
geometrical acceptance, energy and momentum resolutions, and the
efficiency of detectors. Fermi motion~\cite{PARIS} and the off-shell
aspect of the target nucleons inside deuterium~\cite{OFFSHELL}, and
final-state interaction between target and spectator
nucleons~\cite{FSI} were also included in order to describe the $\mmd$
distribution of incoherent events. The shape of the distribution was
found to be sensitive to the off-shell effect implemented. Two methods
of treating off-shell effects for struck nucleons were studied for the
estimation of systematic error~\cite{LEPS_phiD_co}. The individual
yields of coherent and incoherent reactions were extracted by fitting
the $\mmd$ distribution. The results of coherent production are
reported in Ref.~\cite{LEPS_phiD_co}.


\section {Results}

We calculated differential cross sections in the region of
$1.57<E_{\gamma}<2.37$ GeV and $|\tilde{t}|<0.6$ GeV$^2$/c$^2$,
where $\tilde{t}=t-t_{min}$. Here, $t$ is the squared four-momentum
transfer between incident photon and nucleon target, and $t_{min}$
corresponds to the $t$ where the $\phi$ meson is produced at zero
degrees. Because of the deuteron form factor, coherent events appear
mostly in the small $|\tilde{t}|$ region~\cite{LEPS_phiD_co}. The
disentanglement fit using MC simulated coherent and incoherent
components in $\mmd$ spectra was performed in the very forward region
of $|\tilde{t}|<0.4$ GeV$^2$/c$^2$, with the bin sizes of 0.1 GeV for
\Eg and 0.02 GeV$^2$/c$^2$ for $\tilde{t}$. All $\phi$-events with
$|\tilde{t}|>0.4$ GeV$^2$/c$^2$ were counted as incoherent.

\begin{figure}
\vspace{-1cm}
  \includegraphics[width=0.55\textwidth]{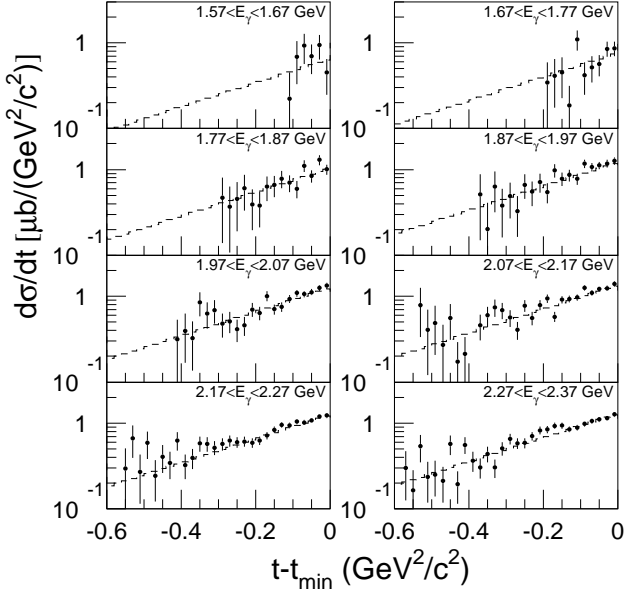}
  \caption{Exponential fit of the \tvarr distribution of incoherent events
    $\gamma d \rightarrow \phi pn$ at the same \Eg bins as
    Fig.~\ref{fig:mmdld2}. The dashed lines are fits of an exponential
    function convoluted with the acceptance. A single slope
    parameter is used for all \Eg bins.}
  \label{fig:tdis_inco}
\end{figure}

With a proper normalization of the photon beam flux, number of target
atoms, tagger efficiency, transporting efficiency and branching ratio
of charged decay of $\phi$ mesons, differential cross sections of
incoherent events $d \sigma / d\tilde{t}$ are displayed in
Fig.~\ref{fig:tdis_inco}. The error bars shown are statistical
only. The differential cross section was fitted with an exponential
function convoluted with MC-estimated acceptance efficiency:
$d\sigma/d\tilde{t}$=$a \exp(b\tilde{t})$, where the fit parameter $a$
is the differential cross section at $\tvarr$=0 ({\it i.e.} zero
degrees) and $b$ is the exponential slope. Within the error range of
the slope $b$ at these \Eg bins, no strong energy dependence is
observed.

\begin{figure}
\vspace{-1cm}
\includegraphics[width=0.55\textwidth]{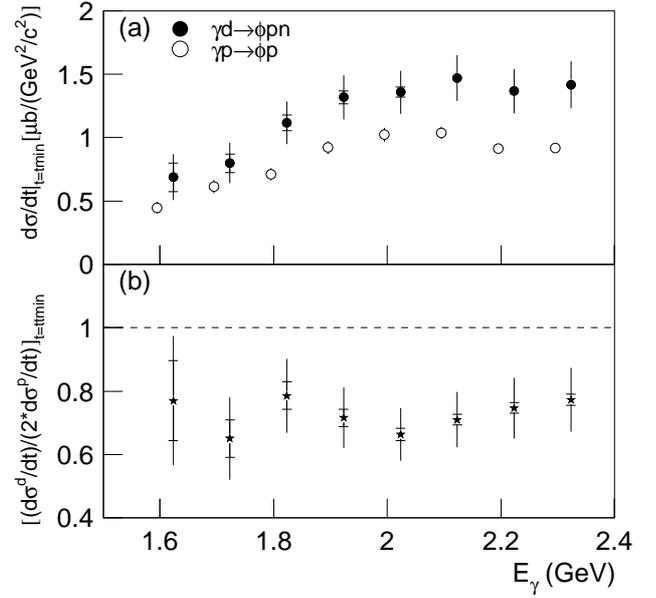}

\caption{(a) The fitted $d \sigma/dt$ at $t=t_{min}$ as a function
of photon energy for incoherent production from deuterium and that from
hydrogen~\cite{LEPS_phi_proton}. The horizontal bars represent the
range of statistical error. The ${\gamma p \rightarrow \phi p}$ data
are shifted by -50 MeV for the clarity of display. (b) Nuclear
transparency ratios of $\phi$-meson photoproduction from deuterium
(A=2) are shown as a function of the photon energy \Eg.}

\label{fig:cross_inco}
\end{figure}

A fit with one common slope parameter for all \Eg bins gives a value
of $b$ to be $3.23 \pm 0.11 ({\rm stat}) \pm 0.16 ({\rm sys})$
c$^2$/GeV$^{2}$ for incoherent events, which is consistent with $3.38
\pm 0.23$ c$^2$/GeV$^{2}$ for production from a free
proton~\cite{LEPS_phi_proton}. The zero-degree cross sections of
incoherent production from deuterium as a function of the photon
energy are shown in Fig.~\ref{fig:cross_inco}(a), together with data
from hydrogen~\cite{LEPS_phi_proton}. The total error is the square
root sum of the statistics and systematic one while the bars represent
the range of statistical errors. The systematic uncertainties arise
from the disentanglement fit (10-15\%), background (5-10\%),
luminosity (5\%) and track reconstruction efficiency (5-10\%).

Since the slope parameters of differential cross sections in the
production from deuterium and hydrogen are rather close, the nuclear
transparency ratios for deuterium, $T_d=\sigma_{d}/(2\sigma_{p})$, can
be evaluated by the ratio of cross sections at zero degrees and are
shown in Fig.~\ref{fig:cross_inco}(b). Compared to the $\phi$
production from a free proton, a significant 25-30\% reduction of the
$\phi$ yield per nucleon is observed for incoherent production from
deuterium. We confirmed that the cross sections of ${\gamma p
\rightarrow \phi p}$ obtained from runs off hydrogen in the present
experiment are in agreement with the reported value in
Ref.~\cite{LEPS_phi_proton}, thus checking the consistency of the beam
and target normalization and the related efficiency factors.

Because the normalization used in the nuclear transparency ratio is
associated with the production from protons, it is straightforward to
speculate that the observed suppression of nuclear transparency ratio
for deuterium stems from a reduction in $\phi$-mesons produced from
neutrons. For example, Titov {\it et al.}~\cite{Titov_iso} suggested
that such a reduction happens because of a destructive interference
effect between unnatural-parity isovector $\pi$ and isoscalar $\eta$
exchange. However it is known that $\phi$-meson photoproduction from
hydrogen is dominated by the processes of natural-parity $t$-channel
exchange~\cite{LEPS_phi_proton}. A possible interference effect in the
unnatural-parity sector is unlikely to induce significant differences
in the cross sections from the proton or the neutron.

\begin{figure}
\vspace{-0.5cm}
\hspace{-0.3cm} 
\begin{minipage}{0.26\textwidth} 
\centering
\includegraphics[scale=0.55]{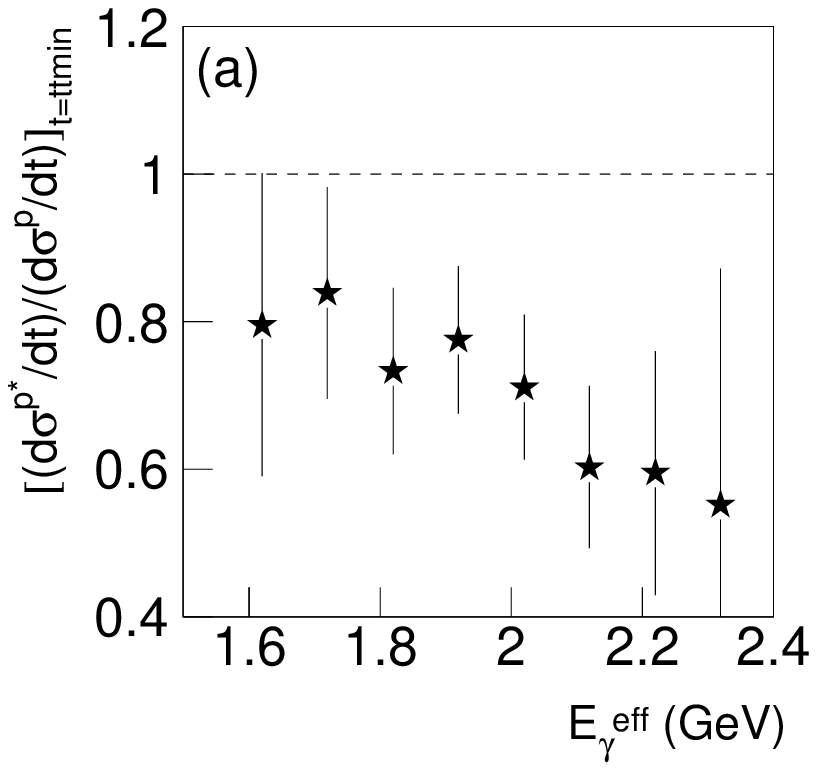}
\end{minipage}
\hspace{-0.6cm} 
\begin{minipage}{0.26\textwidth}
\centering
\includegraphics[scale=0.55]{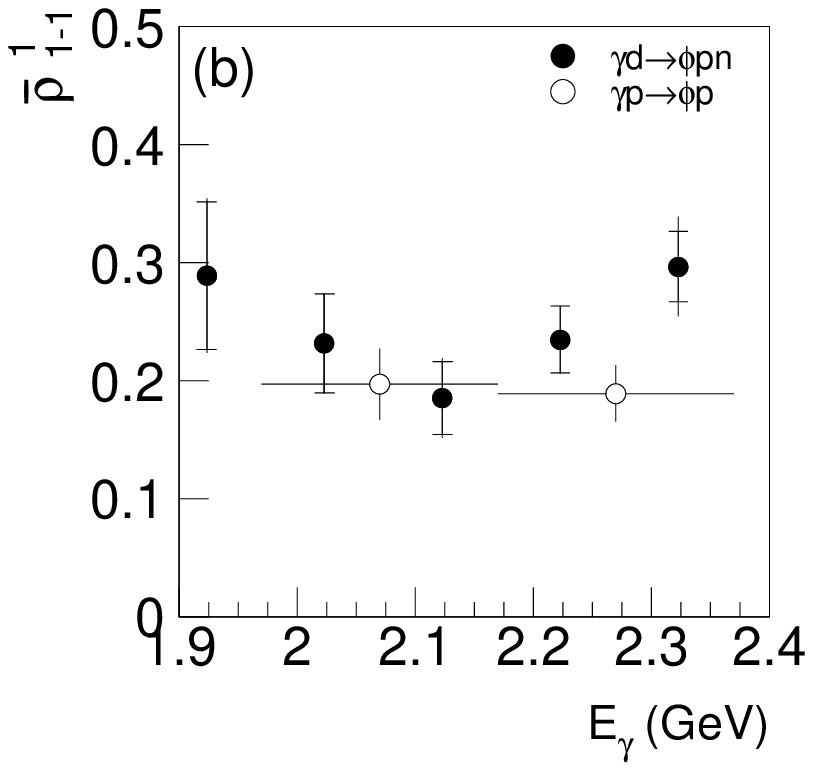}
\end{minipage}

\caption{(a) Ratio of $\phi$-meson photoproduction from the proton
inside deuterium and hydrogen as a function of photon energy \Eg. The
error bars are statistical only. (b) The decay asymmetry
$\bar{\rho}^{1}_{1-1}$ of $\gamma d \rightarrow \phi pn$ inside
deuterium and $\gamma p \rightarrow \phi p$~\cite{LEPS_phi_proton}
as a function of photon energy. The horizontal bars represent the
range of statistical errors.}

\label{fig:rho_inco}
\end{figure}

We examined the isospin dependence of $\phi$ photoproduction by
studying the exclusive $\phi$ events whose final state of a \KK pair
and a proton is fully detected in the spectrometer; the kinematics of
these events is dominated by interactions with the proton inside
deuterium. Though limited by statistics, Fig.~\ref{fig:rho_inco}(a)
clearly shows a similar degree of reduction for quasi-free events from
the proton as compared with inclusive reactions, see
Fig.~\ref{fig:cross_inco}(b). Therefore the reduction in $\phi$ yields
occurs in a similar scale for the incoherent production from both the
proton and the neutron inside deuterium. It also suggests that the
$\pi$-$\eta$ interference effect, which makes a difference in the
production from proton and neutron, is small.


Further information on isospin effects comes from the decay angular
distributions of W($\Phi-\Psi$). These distributions were obtained in
the Gottfried-Jackson frame and in the region of $|\tilde{t}|<0.1$
GeV$^2$/c$^2$ at $1.87<E_{\gamma}<2.37$ GeV. Here, $\Phi$ denotes
the decay azimuthal angles of the $K^+$ in the $\phi$-meson rest
frame.  The azimuthal angle between the photon polarization and the
production plane is $\Psi$. Events with photon energies below 1.9 GeV
were excluded due to insufficient statistics in the angular bins.

The decay angular distribution of W($\Phi-\Psi$) is parametrized as
$1+2P_{\gamma} \bar{\rho}^{1}_{1-1} \cos[2(\Phi-\Psi)]$, where
$P_{\gamma}$ is the polarization degree of the photon
beams~\cite{Titov_spinamp}. In the case of helicity-conservation, the
decay asymmetry $\bar{\rho}^{1}_{1-1}$ reflects the relative
contributions of natural- ($|I_0^N|^2$) and unnatural-parity processes
($|I_0^U|^2$):
$\bar{\rho}^{1}_{1-1}=0.5(|I_0^N|^2-|I_0^U|^2)/(|I_0^N|^2+|I_0^U|^2)$~\cite{Titov_spinamp}. In
two separated regions of missing mass, the decay asymmetry
$\bar{\rho}^{1}_{1-1}$ and the percentage of incoherent events $(R)$
were determined respectively. The individual decay asymmetry of
incoherent ($\bar{\rho}^{1}_{1-1}$$^{\rm inco}$) and coherent
($\bar{\rho}^{1}_{1-1}$$^{\rm co}$) events were extracted from the
difference of measured decay asymmetries and relative weights of two
kinds of events in these two regions, assuming a linear weighting of
each component, $\bar{\rho}^{1}_{1-1}=R\bar{\rho}^{1}_{1-1}$$^{\rm
inco}+(1-R)\bar{\rho}^{1}_{1-1}$$^{\rm
co}$. Figure~\ref{fig:rho_inco}(b) shows the decay asymmetries
$\bar{\rho}^{1}_{1-1}$$^{\rm inco}$ as a function of photon energy at
$E_{\gamma}$=1.87-2.37 GeV. Compared with those for ${\gamma p
\rightarrow \phi p}$ reaction, the decay asymmetries for the ${\gamma
d \rightarrow \phi pn}$ reaction is slightly larger in the region of
$2.17<E_{\gamma}<2.37$ GeV, and agrees well in the region of
$1.97<E_{\gamma}<2.17$ GeV.

Given a value of 0.25 for $\bar{\rho}^{1}_{1-1}$$^{\rm inco}$, the
$\bar{\rho}^{1}_{1-1}$ for the ${\gamma n \rightarrow \phi n}$
interaction would be 0.3 assuming an equal production from either the
proton or the neutron.  One theoretical model gives a compatible
prediction of 0.3-0.35~\cite{Titov_coherent2}. Constrained by the
measurement of decay asymmetries, the possible target isospin
asymmetry is less than 15\% and, the reduction in the $\phi$ yield per
nucleon for incoherent production from deuterium is at most 8\% within
the scenario of interference effect between unnatural-parity parts.
The closeness of decay asymmetries in the interactions with nucleons
and with free protons actually hints at the weakness of the isoscalar
component $\eta$-exchange in the unnatural-parity exchange processes.
This interpretation is supported by a complete dominance of the
natural-parity Pomeron exchange processes in the coherent production,
where the isovector $\pi$-exchange is forbidden~\cite{LEPS_phiD_co}.


\begin{figure}
\vspace{-1cm}
\includegraphics[width=0.55\textwidth]{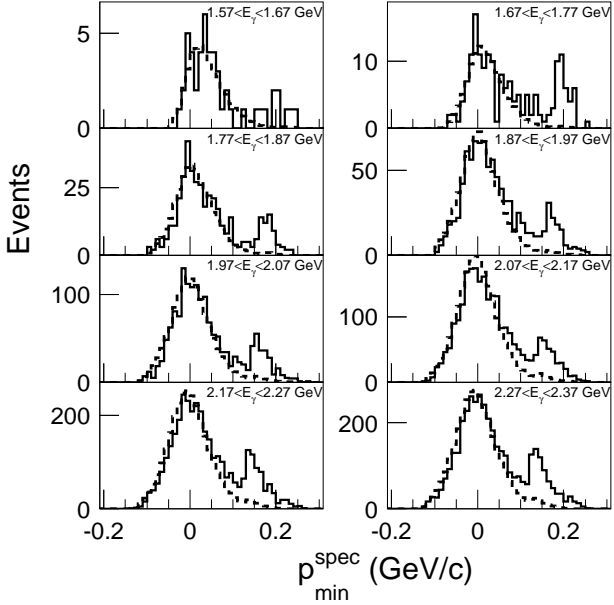}

\caption{The distribution of minimum momentum, $p_{min}$, at
$1.57<E_{\gamma}<2.37$ GeV. The dashed curves are MC-simulated
distributions of incoherent events without final-state interactions.}

\label{fig:qfm}
\end{figure}

Effects of possible final-state interactions and Fermi momentum are
explored by investigating quasi-free incoherent events using the
``minimum momentum spectator approximation'' developed in
Ref.~\cite{Theta_LEPS}. In our acceptance, the Fermi momentum is well
approximated by the so-called ``minimum momentum of spectator'',
$p_{\rm min}^{\rm spec}$, defined as the component of the spectator
momentum in the direction of the momentum of the $np$ pair. The
magnitude of $p_{\rm min}^{\rm spec}$ is about $70/\sqrt{3}$
MeV/$c$. Fig.~\ref{fig:qfm} shows the $p_{\rm min}^{\rm spec}$
distributions for $\phi$ events in eight bins of
$1.57<E_{\gamma}<2.37$ GeV. The main contribution with a peak near
zero comes from incoherent events. Contributions from coherent events
with a deuteron in the final state is characterized by a positive
$p_{\rm min}^{\rm spec}$ around 0.15-0.2 GeV$/c$, which is
approximately equal to half of the momentum of the $np$ pair.

For incoherent $\pi^0$ photoproduction from deuterium, final-state
interactions play a significant role because the final NN state
strongly overlaps with deuteron bound state~\cite{FSI_pi0}. A large
reduction is found in the differential cross section of incoherent
production from deuterons at forward angles, compared to that of
production from free protons. However the situation is rather
different in the case of $\phi$ production; the large mass of the
$\phi$ meson results in a large momentum transfer between the incident
photon and the participant nucleon in the near-threshold production.
The minimum momentum of a rescattered nucleon in the lab system ranges
from 700 MeV/$c$ at threshold to 250 MeV/$c$ at $E_{\gamma}$=2.5
GeV. With such a large momentum for the recoiled nucleon, the
overlapping of final NN state with deuteron bound state is rather
limited. Therefore the effect of final-state interactions is greatly
reduced. As mentioned earlier, the slope parameters of differential
cross sections for deuteron and proton targets in the forward
direction agree well. There also exists an overall agreement between
the distribution of incoherent events and the overlaid curve
representing the MC-simulated distribution of incoherent events
without any final-state interactions shown in Fig.~\ref{fig:qfm}. It
is noted that the discrepancy between the real distribution and the MC
simulation around $p_{\rm min}^{\rm spec}$= 0.1 GeV/$c$ cannot be
fully resolved even with the inclusion of final-state interactions.

Events which are associated with quasifree processes can be selected
with a small $|p_{\rm min}^{\rm spec}|$ value. The nuclear
transparency ratios, for events with the magnitude of $p_{\rm
min}^{\rm spec}$ below 90 MeV, agree well with those obtained using
the disentanglement method. Thus we exclude both final-state
interactions (between participants and spectator nucleons) and Fermi
momentum of target nucleons from the dominating mechanism in
generating the observed reduction.

\section{Discussion}

A target mass number dependence of A$^{-0.28}$ in the nuclear
transparency of $\phi$ mesons was found in the incoherent
photoproduction with nuclei~\cite{LEPS_phiA}. According to this
dependency, the expected nuclear transparency ratio for deuterium is
0.82, which is even slightly above the energy-averaged value
$0.73\pm0.058$, obtained from the ratios shown in
Fig.~\ref{fig:cross_inco}(b). Since the deuteron is composed of a
loosely bound proton and neutron, where the nuclear medium effect is
minimal, the present observations strongly suggest that some effect
other than nuclear density is necessary to achieve a complete
understanding the reduction of $\phi$ production in the nuclear
medium~\cite{phiA_oset}. For example, since the present results are
obtained near the production threshold, the momenta of produced $\phi$
mesons are relatively small in the range of 1-2 GeV/c. The $\phi$
meson, which could fluctuate to either a \KK pair or an $\omega$
meson~\cite{phiA_sibirtsev}, is likely to have a larger cross section
with nucleons nearby. In the low-energy region, a two-step processes
and/or coupled-channel effects might induce a more significant loss of
$\phi$-mesons in nuclei than the current theoretical estimates.

\section{Summary}
In summary, differential cross sections and decay asymmetries of
incoherent $\phi$-meson photoproduction from deuterons were measured
at forward scattering angles near threshold. In comparison with those
from proton, the decay asymmetry is similar, but the production cross
section per nucleon shows a significant reduction. The reduction is
common in incoherent production from both the proton and the neutron
inside deuterium. The target isospin asymmetry is found to be small
and cannot account for the large suppression of the nuclear
transparency ratio for the deuteron. The present work suggests that
the nuclear transparency ratio of $\phi$-meson photoproduction is
sensitive to the details of the nuclear structure. It should also be
an important baseline to differentiate nuclear density
effects. Further theoretical work on refining the $\phi$-N interaction
within a nucleus is required to explain such a large reduction near
threshold from deuterium. Hence the existence of nuclear medium
effects, or any other interesting mechanisms, can be identified in a
future study of $\phi$-meson production from nuclei.


\section*{Acknowledgments}

The authors thank the contributions of SPring-8 staff in the operation
of the LEPS experiment. This research was supported in part by the
Ministry of Education, Science, Sports and Culture of Japan, by the
National Science Council of Republic of China (Taiwan), Korea Research
Foundation Grant (2006-312-C00507) and National Science Foundation
(NSF Award PHY-0244999).



\end{document}